\documentclass{elsart}

\input epsf

\def\plotone#1{\centering \leavevmode
\epsfxsize=\columnwidth \epsfbox{#1}}

\newcommand{\rh}{r_{\bullet}}
\newcommand{\msigma}{$M_{\bullet}$--$\sigma_*$\ }
\newcommand{\msun}{M_{\odot }}

\newcommand{\kms}{\, {\rm km \, s}^{-1} }

\newcommand{\pc}{\, {\rm pc} }

\newcommand{\yr}{{\, \rm yr} }
\newcommand{\Gyr}{{\, \rm Gyr} }

\begin{document}

\begin{frontmatter}
\title{Feeding black holes at galactic centres by capture from isothermal cusps}
\author{HongSheng Zhao\thanksref{a1}}
\author{Martin G. Haehnelt\thanksref{a1}}
\author{Martin J. Rees\thanksref{a1}}
\thanks[a1]{Institute of Astronomy, Madingley Road, Cambridge CB30HA, England}

\begin{abstract}
The capture rates of stars and dark particles onto supermassive black
holes depend strongly on the spatial and kinematical distribution of
the stellar and dark matter at the centre of bulges and elliptical
galaxies.  We here explore the possibility that all ellipticals/bulges
have initially isothermal cusps ($\rho \propto r^{-2}$).  If the
orbits can be adequately randomized a significant fraction of the
total mass of black holes in the bulges of galaxies will be due to the
capture of stars and dark matter.  The dark matter fraction of the
total mass captured may be as high as 20--40 percent for typical cold
dark matter halos.  A tight relation $M_{\bullet} \sim 10^8
(\sigma_*/200 \kms)^5$ between black hole mass and stellar velocity
dispersion can arise at the high mass end ($M_{\bullet} \ge 10^8
\msun$) if these giant black holes grow primarily by the capture of
stars without tidal disruption.  For smaller black holes a shallower
$M_{\bullet}-\sigma_*$ relation with larger scatter is
expected. Efficient randomization of the orbits can be due to remnant
accretion discs or the dense central regions of infalling satellites
which can avoid tidal disruption and sink to the sphere of influence
by dynamical friction. The presence of an isothermal cusp and the
reduction of the relaxation time scale at the sphere of influence
enhance the estimated tidal disruption rate of stars to $\sim 10^{-4}$
-- $10^{-2}\yr^{-1}$ per galaxy.  Disruption flares in bright galaxies
may thus be as frequent as a few percent of the supernovae rate at
moderate redshifts when the galaxies still had an isothermal cusp. The
efficient replenishment of the loss cone also explains why the
supermassive binary black holes expected in hierarchically merging
galaxies do generally coalesce as suggested by the observed relation
between black hole mass and the inferred mass of stars ejected from an
isothermal cusp.
\end{abstract}

\end{frontmatter}

\section{Introduction}

The confidence in the measurement of the masses of supermassive black
holes in nearby galaxies has significantly increased in the  last couple of
years (see Merritt \& Ferrarese 2001 for a recent review). This is
mainly due to the newly established  correlation between black 
hole mass and velocity dispersion of the bulge of the host galaxy
(Gebhardt et al. 2000, Ferrarese \& Merritt 2000). Most if not all
galactic bulges appear to contain a black hole with mass 
$M_{\bullet} \propto  \sigma_*^{4-5}$.  For the published samples of 
reliable black hole masses this correlation  appears much tighter 
than that between  bulge mass and bulge luminosity  
(Kormendy \& Richstone 1995, Magorrian et al.~1998).  
A number of suggestions have 
been made which can  explain the slope of the correlation 
(Silk \& Rees 1998; Haehnelt, Natarajan \& Rees 1998; Kauffmann \& Haehnelt
2000; Haehnelt\& Kauffmann 2000a; Ostriker 2000;
Burkert \& Silk 2001;  Adams, Graf \& Richstone 2001) 
but little has  been offered to
explain its apparent tightness. The physical processes invoked
to regulate the black hole mass depend generally on the conditions close to 
the black hole at radii much smaller than those at which the 
velocity dispersion of stars is measured. Moreover, the scatter 
in correlations of observed properties of galaxies like the 
Tully-Fisher and Faber-Jackson relation is  much larger. 
Haehnelt\& Kauffmann (2000b) demonstrated that their model is consistent 
with the observed scatter. However, the small scatter  does  not 
occur naturally in such a model where galaxies build up by
hierarchical merging. Should the tightness of the correlation 
between black hole mass and bulge velocity dispersion stand 
the test of time a physical mechanism which links the black hole mass 
and the velocity dispersion of the stars in the bulge more directly
may be required.  Capture of stars by the
black hole for instance is a process that depends straightforwardly 
on the stellar velocity dispersion (Rees 1988). 
The main problem is that the orbits of stars with 
sufficiently low angular momentum are 
generally assumed to be rapidly depleted, inhibiting efficient 
growth  by accretion of stars (Sigurdsson \& Rees 1997,
Magorrian \& Tremaine 1999, Syer \& Ulmer 1999).  
The density profiles of bright ellipticals exhibit pronounced 
breaks within which the density profile becomes significantly 
shallower than isothermal at  radii of a few hundred
parsecs (Gebhardt 1996).  Very little feeding of the black hole 
can come from stars at large radii, 
where the loss cone becomes prohibitively small. 
These shallower cores may however have formed very 
recently due to the ejection of stars by the supermassive 
binary black holes (Milosavljevi\'c \& Merritt  2001, 
Ravindranath, Ho \& Filipenko 2002)
expected  in hierarchical merging galaxies (Kauffmann \& Haehnelt
2000).  Strong supports for this idea come from the observed correlation  
between the mass of the black hole and the mass inferred to be ejected 
if the galaxy started out with a cusp with a density profile close 
to isothermal.
This makes it likely that bulges form with a stellar density 
distribution which is close to isothermal all the way down to the 
sphere of influence of the supermassive black hole.  The number of
stars on low-angular momentum orbits and thus the rate of capture of 
stars by the supermassive  at the centre of nearby galaxies must
have been much larger prior to the destruction of the isothermal cusp.
Here we explore this idea in more detail.

\section{Growing supermassive black holes by the capture of stars}

\subsection{Direct capture and tidal disruption} 

For supermassive black holes more massive than  
\begin{equation}
M_{\rm disr } \approx  10^{8} \msun
\end{equation} 
direct capture of solar-type stars is  
possible if the  angular momentum of 
the star is smaller than some critical value (Frank \& Rees 1976). 
For a Schwarzschild black hole this value is given by 
\begin{equation}
J_{\rm cap} \le  \frac{l G M_{\bullet}}{c}, \qquad l=4;
\end{equation} 
for a Kerr BH, $l$ is slightly $>4$ 
for an incoming particle retrograde to the spin of the hole,
and slightly $<4$ for a direct particle.
In the case of less massive black holes solar-type stars  with
angular momentum smaller than $ \sqrt {2 G M_{\bullet} r_{\rm disr}}$ 
will be tidally disrupted before they reach the horizon at a radius 
\begin{equation}
r_{\rm disr} =  \left ( \frac {M_{\bullet}}{M_{\rm disr}} \right )^{-2/3} 
r_{\rm S},
\end{equation}
where $r_{\rm S} = 2GM_{\bullet}/c^2$  denotes the Schwarzschild
radius. The ``loss cone''  of stars with sufficiently small angular 
momentum is larger than that for direct capture by a 
factor  $(M_{\bullet}/M_{\rm disr})^{-2/3}$. Which fraction 
of the debris can be accreted  by the black hole is uncertain 
and depends on how much gas is expelled by a possible radiation-driven
wind (Rees 1988).  Dark matter   particles 
with angular momentum as low as given in equation (2)  
will also be accreted by direct capture. We will discuss  
this in more detail in section 3.4.

\subsection{A simple argument for the \msigma relation}  
\begin{figure}
\plotone{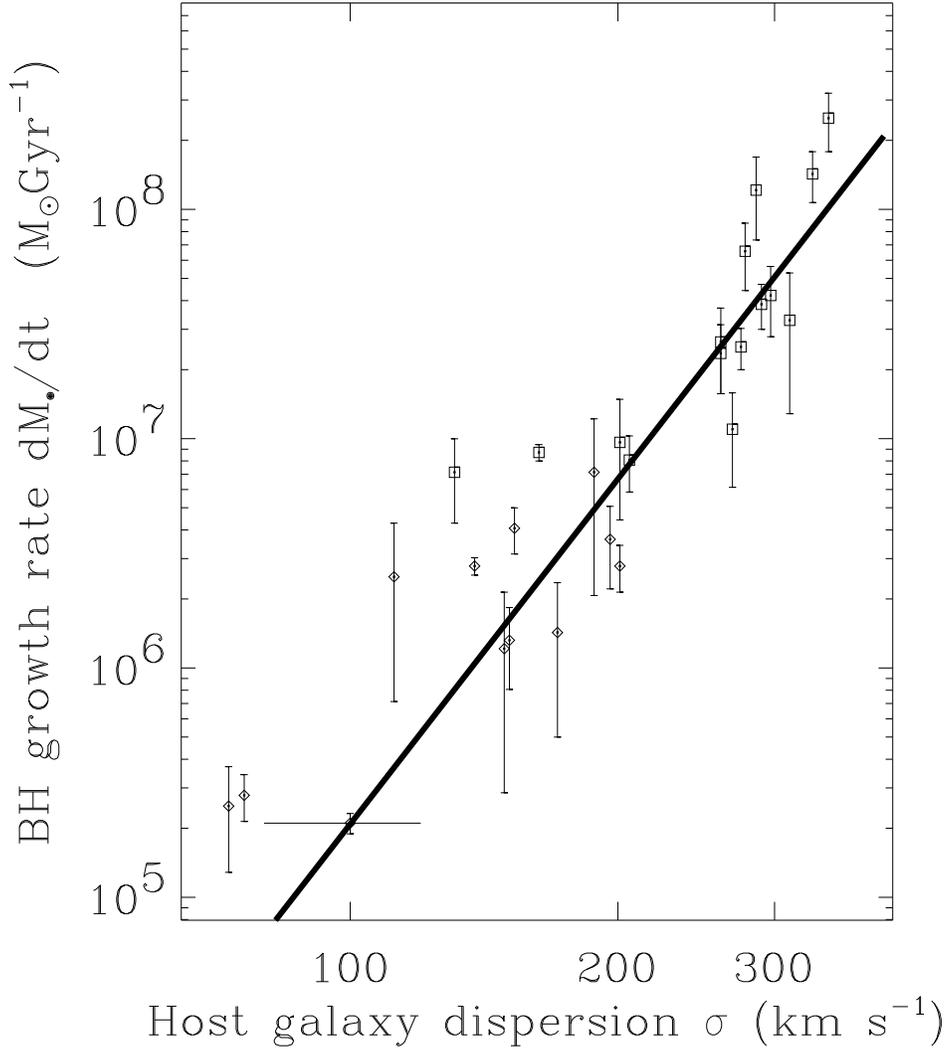}
\label{fig:grow.ps}
\caption{The rate $\dot M_{\rm cap} (\rh)$ for the capture of stars by
the black hole as given by equation (8) compared to the growth rate of 
an observed sample of black holes (Ferrarese \& Merritt 2000) 
time-averaged over a Hubble time of 14 Gyr.}
\end{figure}

Let us first consider the case of supermassive black 
holes more massive  than $M_{\rm disr}$ embedded in an 
isothermal stellar cusp ($\rho \propto r^{-2}$).  
For an  isotropic Gaussian velocity 
distribution  a fraction 
\begin{equation}
f_{\rm cap} (r) = \left (\frac {J_{\rm cap}}{2 \sigma_* r}\right )^2 =   
\frac {2 r_{\rm S} \rh} {r^2} = 4\,  \frac {\sigma_*^2} {c^2} 
\left (\frac {\rh} {r}\right )^2 
\end{equation} 
of the stars at radius $r$ has sufficiently low angular momentum 
and will be captured within a dynamical time.  
Here we have defined the radius of the sphere of influence
\begin{equation}
\rh= \frac {G M_{\bullet}}{\sigma_*^2} 
\approx  15  \left ( \frac{\sigma_*}{200\kms}\right )^{2.8}   \pc.  
\end{equation}
For the right hand side of equation (4) we have assumed 
the observed relation 
\begin{equation}
M_{\bullet} =1.4\times 10^{8} \msun \, (\sigma_*/200 \kms)^{4.8}
\end{equation} 
as quoted by Ferrarese \& Merritt (2000). 

For a (singular) isothermal cusp we can now write  
the rate at which the ``loss cone'' of stars with low angular 
momentum  is cleared as a mass accretion rate. of stars 
captured from  radius r. Outside $\rh$ this is given by
\begin{eqnarray}
\dot M_{\rm cap}(r) &=& 
4 \pi \, \rho (r)\,  r^2 \sigma_* \, f_{\rm cap} \\\label{rateatrh}
&=& 8 \, \frac{\sigma_*^3}{G} \, \frac {\sigma_*^2}{c^2}\,  \left 
(\frac{\rh}{r} \right )^2. 
\end{eqnarray}

There are a few things to note. 
First the dynamical time at the sphere of 
influence is  only $t_{\rm dyn} = r/\sigma_* = 
7\times 10^{4}(\sigma_*/200\kms)^{1.8}$ yr. 
To achieve significant growth the ``loss cone'' has to 
be refilled many ($\sim f_{\rm cap}^{-1}$) times. 
We will come back 
to this point later. Second the accretion rate 
at  $\rh$ does not depend on mass and 
decreases as $r^{-2}$ at larger radii. 
A  black hole growing  by  capture of stars from 
an isothermal cusp will grow  inside-out from the sphere 
of influence.  Due to the strong dependence on radius the mass 
accretion rate is very sensitive to the stellar distribution just 
outside  the sphere of influence of the black hole. 
In Figure 1 we compare the accretion rate given by equation~(\ref{rateatrh})
with those which would be necessary to grow the black holes 
in observed galaxies over a Hubble time. 

The growth rate will be dominated by capture of stars 
from $r\sim \rh$ and within a time $t_{0}$  the black hole  
grows  to a mass 
\begin{equation} M_{\bullet} 
\approx 10^8 \msun \left (\frac {\sigma_*}{200 \kms} \right)^5 
\left (\frac {t_0} {14 \Gyr}\right). 
\end{equation}
if the loss cone stays full. 
Slope {\it and} normalization of the observed \msigma relation 
can thus be explained by capture of stars if the isothermal cusp
existed and the loss cone can be kept filled for a fair fraction of the 
Hubble time. The  velocity dispersion is independent of radius 
in isothermal stellar systems. The scatter in the  observed 
correlation would then be reduced to that in $t_{0}$.  

Dynamical relaxation makes stars wander in  angular momentum
space, and fill the loss cone on a timescale 
$t_{\rm fill}\sim  f_{\rm cap} t_{\rm relax}$ 
(e.g. Magorrian \&Tremaine 1999, Syer \& Ulmer 1999).
If  $t_{\rm fill} >t_{\rm dyn}$ the mass accretion would be controlled 
by the diffusion rate of stars into the loss cone. The mass accretion 
rate would be smaller by a factor $t_{\rm dyn}/t_{\rm fill}$ and  
the mass would be smaller by a factor $t_{0}/t_{\rm relax}$.

\subsection{Why loss cone depletion may not be important} 

Previous estimates of capture  rates of solar-type stars 
by supermassive black holes are of the order  of $10^{-6}$ -- $10^{-4} 
\yr^{-1}$  (Magorrian \& Tremaine 1999, Syer \& Ulmer 1999)
much lower than required for  significant growth of the 
supermassive black holes in nearby galaxies. These  rather low capture rates 
result mainly from the following  assumptions: 
\begin{itemize} 
\item 
the supermassive black holes are fed from 
shallow  cores typical for bright 
present-day elliptical 
galaxies; 

\item the distribution of the stars feeding the supermassive back hole
is spherical or axisymmetric and angular momentum is conserved; 

\item star-star two body-relaxation is the dominant 
process for repopulating the loss cone.   

\end{itemize}

However, hierarchical structure formation scenarios 
suggest that present-day galaxies have evolved 
strongly with redshift. The first two assumptions are thus most likely 
not justified for the major fraction of the past 
history of present-day galaxies. We have already discussed that 
the shallow cores in bright galaxies may have been established very 
recently due to ejection of stars by supermassive binary black
holes and that all ellipticals/bulges  may initially have had a
cusp with an isothermal density distribution down to the 
sphere of influence of the black hole. 

The chaotic motions induced by the presence of a black hole have been
demonstrated to destroy the triaxiality of galaxies outside the sphere 
of influence out to about a hundred times $\rh$ in a few crossing times 
(Gerhard \& Binney 1985, Merritt \& Quinlan 1998, Sellwood 2001). 
Poon \& Merritt (2002), however, have argued that at 
smaller radii close to the sphere of influence  stable triaxial nuclei 
may persist. 
This brings into question to what extent angular momentum 
in an isothermal cusp around a supermassive black hole would be 
conserved and whether refilling of the loss cone may occur faster
than on a star-star relaxation timescale ({\it cf} Holley-Bockelmann 
et al 2002).

For the observed \msigma relation  the timescale 
for the relaxation of the energy of stellar orbits 
by star-star interactions  at the sphere of influence can 
be written as 
\begin{equation} 
t_{\rm relax}(\rh) = \frac {N_*}{8\ln N_*} \,t_{\rm dyn}(\rh)
\approx 10 \left (\frac{\sigma_*}{150\kms}\right )^{6.6} \Gyr,
\end{equation} 
where $N_*$ and $\ln N_*$ denote the number of stars and the 
relevant ``Coulomb  logarithm'', respectively (Binney \& Tremaine
1987).  This  relaxation time is longer than a Hubble 
time for galaxies with $\sigma_{*}> 150 \kms$, so the amount of black 
hole growth would be reduced by a factor $(\sigma_*/150\kms)^{6.6}$. 

The two-body relaxation time scale is also strongly dependent on the 
mass of the perturbing objects. The presence of a single heavy 
perturber of mass $m_{\rm pert}$ in a  system of $N_*$ stars will 
reduce the star-star energy 
relaxation time scale by a factor $\sim m_{\rm pert}^2 /N_* m_*^2$  
(see also Polnarev \& Rees 1994). At the  sphere of influence  
$N_* m_* \sim M_{\bullet}$  and  the two-body energy relaxation time-scale 
for an isothermal 
cusp  is reduced  below 10 Gyr for mass ratios 
\begin{equation}
\frac{m_{\rm pert}}{m_{\bullet}} \ge 0.0002\, \left({\sigma_* \over 200\kms}\right)^{0.9}. 
\end{equation}
A population of smaller mass perturbers with mass 
fraction $f_{\rm pert}$ will reduce the relaxation time-scale 
by a factor  $f_{\rm pert} \, (m_{\rm pert}/ \msun)$   
and the two-body energy relaxation time-scale  at the sphere influence 
is reduced  below 10 Gyr for a mass fraction  
\begin{equation}
f_{\rm pert} \ge 0.007\, \left(\frac{m_{\rm pert}}{1000\msun}
\right)^{-1}\left (\frac {\sigma_*}{200\kms}\right)^{6.6}. 
\end{equation}

Note, however, the important process for our models is the relaxation of the 
angular momentum not energy since
stars are captured from the sphere of influence directly.
The relaxation of angular momentum is 
significantly easier to achieve than that of energy and the above timescales 
should be substantially shorter in the presence 
of a resonant perturber or a small number of resonant perturbers 
(Rauch \& Tremaine 1996).

\subsection{Resonant angular momentum relaxation due to remnant 
accretion discs}

Typical supermassive black holes  have accreted a 
substantial fraction of their mass via an accretion disc
The phases of rapid accretion are
probably short but when the mass supply from larger radii drains mass
accretion rates will drop and the viscous time scales will
increase (e.g. Haehnelt \& Natarajan \& Rees 1998). Phases of 
rapid accretion should thus leave remnant  accretion discs behind.  
The disc should typically contain  molecular gas at a temperature of
100K.  For an $\alpha$-disc the scale height at the sphere of
influence  would be  about $3\times 10^{-3}$ the radius and the 
viscous accretion time scales would be $\alpha^{-1} (H/R)^{-2} 
\sim \alpha^{-1}10^{5}$ times the dynamical time.  For a $10^{8} \msun$ 
black hole this is longer than the Hubble time for $\alpha
\sim 0.3$. For small accretion rates $\alpha$ should be much smaller
and for more massive black holes the numbers become even larger. It
is thus safe to assume that in phases of low accretion from larger
radii remnant accretion discs should survive for a Hubble time. 
Plausible disc to hole mass ratios would be in the range $10^{-3}$
to $10^{-2}$. Note that a massive disc  will be little affected by the 
drag due to dynamical friction. 

The gas in the remnant disc will move on Keplerian orbits in the spherical
potential of the black hole. It will act as a massive resonant
perturber akin to the collective resonant relaxation due to stars on
resonant orbits described by Rauch and Tremaine (1996) and will reduce
the time scale for angular momentum relaxation of stars at the sphere
of influence to well below a Hubble time.  Note again that we do not 
need to achieve energy relaxation as we have assumed direct 
capture of stars from the sphere of influence. This is 
different from the case discussed by Rauch and Tremaine who remarked 
that loss cone refilling will not be enhanced if energy relaxation is 
required for stars to migrate to smaller radii. Unlike the time scale 
for angular momentum relaxation the time scale for energy relaxation
is not affected by resonant orbits.

\subsection{Perturbing galactic nuclei}

The centres of galaxies  building  up by hierarchical merging 
are continuously perturbed by a sequence of infalling smaller
galaxies. E.g. in the model of Kauffmann et al.~ (2002) 
bright galaxies have typically undergone a  
few mergers with galaxies  of 10\% their mass 
and about 10-20 mergers with galaxies 1\% their mass
since $z\sim 1$.   
The outer parts of the infalling galaxy will be progressively 
tidally stripped while it sinks
to the centre on a dynamical friction  time scale.  
The mass of the remnant surviving tidal stripping  will be 
$m_{\rm pert} \sim (\sigma_{\rm sat}/\sigma_{*})^3\,(r/\rh)\,m_{\bullet}$
where we have assumed that the density profile 
of the satellite is also isothermal.
The resulting relaxation time scales as 
$( \sigma_{\rm sat}/\sigma_*)^{-6}$ and   
at radius r satellites with stellar velocity dispersion 
\begin{equation}
\sigma_{\rm sat} \ge 0.06 \,  \left(r \over \rh\right) ^{1 \over 6} 
\left (\frac {\sigma_*}{200\kms}\right )^{0.3}\, \sigma_{*}
\end{equation} 
reduce the stellar relaxation timescale below 10 Gyr. A significant
fraction of the total mass of stars in the cusp may actually be made  
up  from the debris of such tidally disrupted galaxies which   
then can keep the low-angular momentum orbits continuously 
populated.   The streamers left over from the tidal disruption 
of the outer parts of the infalling satellite galaxies  will 
also contribute to the reduction of the relaxation timescale 
(Tremaine \& Ostriker 1999). The massive black holes 
with masses intermediate between those of stellar and 
supermassive black holes which may be left over 
from a pre-galactic episode of star formation 
would be a plausible population of more numerous less 
massive perturbers (Madau \& Rees 2001). The dynamical 
friction time scale would be short enough for 
$1000 \msun$ black holes to sink to the sphere of influence 
from about 10 $\rh$. A fraction of $10^{-3}$ of the 
total mass in the bulge in such intermediate mass black 
holes would thus also be sufficient to reduce the energy relaxation 
timescale below a Hubble time. A smaller fraction would be sufficient 
to reduce the angular momentum relaxation time scale. 
Note that these black holes 
are more easily captured into close bound orbits by 
the central supermassive black holes than stars 
(Sigurdsson \& Rees 1997). Intermediate mass black holes in 
close orbits around supermassive black holes 
with $ M_{\bullet} \le 3 \times 10^{6} \msun$ 
are good candidates for a detection by LISA (Madau\& Rees 2001).

\section{Further implications and potential problems}

\subsection{Other accretion modes} 

Capture of stars is not the only growth mechanism of black 
holes.  Haehnelt, Natarajan \& Rees (1998)  have argued (see also  Haehnelt 
\& Kauffmann 2001) that the black hole mass density inferred by optical
bright accretion is of the same order but somewhat lower than the
total mass density inferred from nearby galaxies. This  leaves some  
room for other modes of accretion.
There are, however, large uncertainties in both the estimates of the 
black hole mass density necessary to produce the QSO emission 
and the total mass density of remnant black holes.  
Merritt and Ferrarese (2001) e.g. conclude that optical 
bright accretion in QSOs could account for the total 
observed black hole mass density in nearby galaxies. Probably
at least 30 percent of the total mass density of supermassive 
black holes in nearby galaxies is already accounted for by accretion of gas
(mostly at $z\ge 1.5$). Accretion of gas where 
the optical emission is obscured by dust will further 
increase this number (Fabian et al. 1998, Haehnelt et al. 1998). 
Rather than establishing the \msigma relation 
the capture of stars may thus just tighten and 
refine the exact slope of a pre-existing rough correlation between
black hole mass and  bulge properties. 

\subsection{Tidal disruption of stars} 

So far we have neglected that  black holes smaller than 
$m_{\rm disr} \sim 10^{8} \msun$ can tidally disrupt stars before 
they reach the horizon (Rees 1988). This  complicates matters  as it 
increases the loss cone by about a factor 
$(M_{\bullet}/M_{\rm disr})^{-2/3}$. The increased size of the loss 
cone should lead to a flattening of 
the \msigma relation at small masses but probably not as much 
as to $M\propto \sigma_*^{3}$. It will be increasingly difficult to 
keep the enhanced loss cone continuously filled at smaller masses 
for a Hubble time. Furthermore, hierarchical 
galaxy formation predicts the merger rates 
to decrease rapidly with decreasing bulge luminosity
(Kauffmann et al. 2002).  The scatter 
of the \msigma relation is therefore expected to increase at smaller masses. 
The number of  black hole mass determinations in fainter galaxies is  
small and observational constraints  are still weak. We further note 
that if loss cone refilling is  as efficient as assumed here then 
the rates for flares due to the tidal disruption of stars in bright
galaxies  will  be significantly  larger at moderate redshift than nearby 
 (Syer \& Ulmer 1999, Magorrian \& Tremaine 1999). 
If the mass of tidally disrupted stars is a fraction $f_{\rm disr}$  of
the black hole mass then the corresponding rate of flares 
should be on average a fraction $\sim 0.2 f_{\rm disr}$ of 
the  supernova rate if we assume one supernova 
per $100\msun$ in stars and $M_{\bullet}/M_{*} \sim 0.002$
The peak-luminosities of the flares can  reach the Eddington 
luminosity of the black hole and are likely to  significantly 
exceed those of supernovae. 
Searches for  high-redshift supernovae may thus detect these flares 
in appreciable numbers.

\subsection{Supermassive binary black holes in hierarchically merging galaxies}

The black hole hosted by  the infalling satellites  
will form a supermassive binary with the primary black hole. 
When the circular velocity of the binary equals  
$\sim \sigma_*$
the binary will harden   
due to scattering of stars 
similar to the hardening of binary stars in star clusters. 
If the binary separation can shrink to a separation
\begin{eqnarray}
a_{\rm gr} &\sim& 4 \times 10^{3} 
\left (\frac{m_1+m_2}{10^8\msun}\right )^{-1/4}
\left ( \frac{m_2}{m_1}\right)^{1/4}  r_{\rm S} \nonumber\\
&&\nonumber\\
&\sim& 
4\times 10^{-2} \left ( \frac{\sigma_*}{200\kms}\right)^{3.6}      
\left ( \frac{m_2}{m_1}\right)^{1/4}     
\pc \nonumber \\
\end{eqnarray}
the supermassive black holes will spiral together within 10 
Gyr due to the emission 
of gravitational waves (Peters 1964). 
At a separation $a_{\rm gr}$ 
the loss cone for binary hardening 
is larger by a factor 
$\sim 4 \times 10^{3}  (\sigma_*/200\kms)^{-1.2}
(m_1/m_2)^{-1/4} $
than that for direct capture of stars. The evolution time scale is, 
however, again set by the stellar relaxation time scale on which 
low-angular momentum orbits are repopulated.  
Whether supermassive binary black holes actually reach this separation 
in normal galaxies has  been a  matter of debate. 
Begelman, Blandford, \& Rees (1980) have argued that in typical bright 
elliptical galaxies the timescale is longer than a Hubble time 
and that accretion of gas would be required to reduce the 
separation sufficiently for gravitational radiation 
to become important if loss cone depletion occurs. 
Yu (2002) in a detailed analysis  
of a sample of nearby galaxies has shown that the supply of
low-angular momentum stars is sufficient 
to reach $a_{\rm gr}$ if the galaxies are  significantly 
flattened or triaxial.  As discussed above,  for an isothermal cusp 
the problem is strongly alleviated (see also 
Milosavljevi\'c \& Merritt 2001 and Gould \& Rix 2000) 
and the presence of the dense remnants of infalling satellite galaxies 
can reduce the stellar relaxation timescale 
at the sphere of influence, and thus also the coalescence timescale of 
supermassive binary black holes in bright galaxies, 
well below a Hubble time. This fits in well  
with  the observed correlation of inferred ejected core mass and
black hole mass (Milosavljevi\'c \& Merritt 2001, 
Ravindranath et al. 2001).  

\subsection{Accretion of dark matter} 
\begin{figure}
\plotone{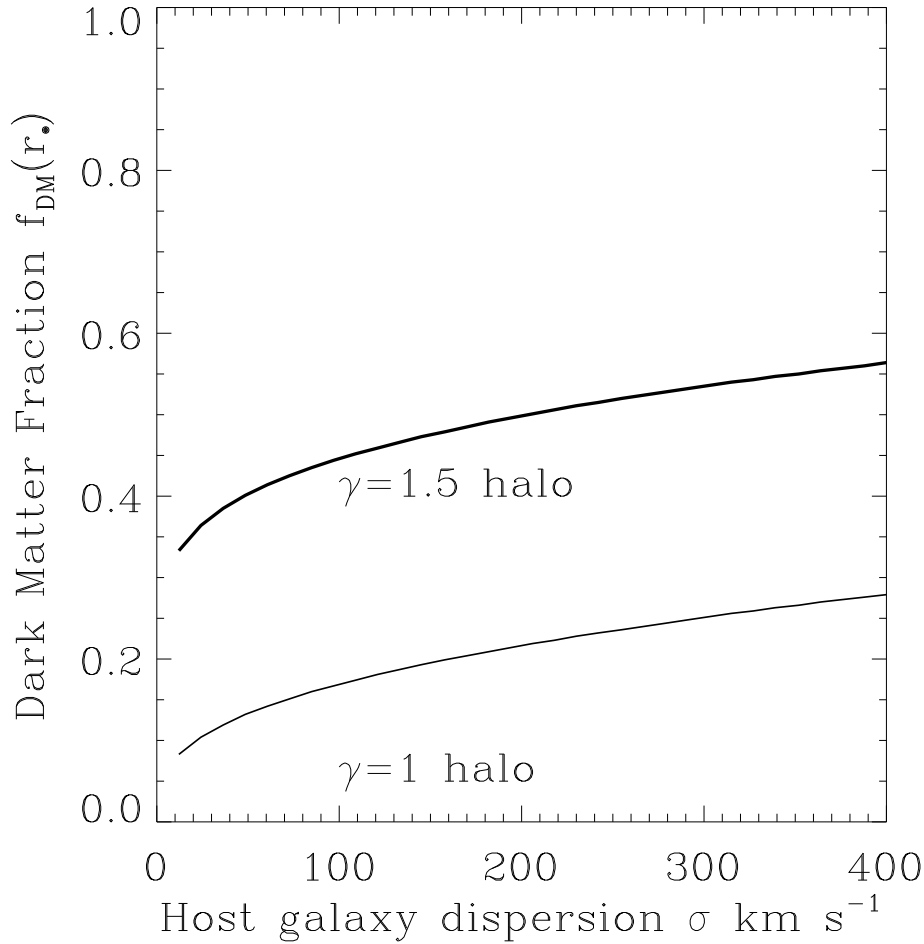}
\label{fig:dm.ps}
\caption{Dark Matter fraction near the sphere of influence of the
black hole. The lower thin line  assumes
an initial dark matter halo density profile as given by 
Navarro, Frenk \& White (1996) with a $\rho_{\rm DM} \sim r^{-1}$ cusp,
and the upper thick line  assumes a density profile as given by  
Moore et al. (1998) with a $\rho \sim r^{-1.5}$ cusp.
The halo concentration parameter and 
the galaxy dispersion are assumed to scale with 
the halo mass as given in Klypin et al. (2001).}
\end{figure}

There is good evidence that in the outer parts of 
bright ellipticals/bulges the gravitational force is 
dominated by  dark matter halos (see Gerhard  et al.~2001
for a recent study). However, the mass fraction of dark matter 
particles in the core of these galaxies is somewhat uncertain 
and will depend on the density profiles of the dark matter 
halo and the baryons in the galaxy. The density profiles of
stellar nuclei, bulges and dark matter halos can be 
well described by a double power-law of the form $\rho \propto 
\tilde{r}^{-\gamma}(1+\tilde{r}^\alpha)^{\gamma-\beta \over \alpha}$, 
where $\tilde{r}$ is a rescaled radius, while  $\gamma$, $\beta$ and $\alpha$ 
prescribe the steepness of the inner cusp, the outer slope and 
the sharpness of the transition, respectively (Zhao 1996). 
For CDM-type structure formation models the dark matter has a 
cusp  $\rho_{\rm DM} \propto r^{-\gamma}$ with $\gamma$ in the 
range 1-1.5 (Navarro, Frenk \& White 1996; Moore et al.~1998, Klypin et
al.~2001). These will, however, be significantly 
modified when the baryons concentrate at the centre 
and dominate the gravitational force. This effect  
can be modelled  by adiabatic contraction.  If 
the final potential is isothermal outside $\rh$ 
due to the formation of the stellar nucleus and bulge
the dark matter will attain a  power-law distribution with 
slope $ \gamma' = (6-\gamma)/(4-\gamma) = (1.8,1.66)$, 
for $\gamma =(1.5, 1)$,  respectively.
The dark matter fraction of the total mass inside $\rh$ 
available for capture is given by 
\begin{equation}
f_{\rm DM} (\rh) =
{M_{\rm DM}(<\rh) \over M_{\rm DM}(<\rh)+ M_{*}(<\rh)}
\sim \left(\rh \over r_{\rm G}\right)^{2-\gamma'},
\end{equation}
where $r_{\rm G}$ is the radius outside which
the dark matter dominates the stars.  The fraction $f_{\rm DM}$ 
is of order 20 percent for $\gamma =1.5$ and  
is rather insensitive to $r_{\rm G}$ for a reasonable range 
of 5-50 kpc. 

Figure 2 shows  the dark matter fraction at 
$\rh$ for an isothermal stellar distribution adiabatically 
grown within more realistic dark matter halo profiles
(see Ullio, Zhao \& Kamionkowski (2001) for details of the 
calculation). The values of the dark matter fraction are in the 
range of 20-50 percent. 
The upper end is probably already too large to be consistent 
with the mass-to-light ratios in elliptical galaxies
(Gerhard et al. 2001).  Note that the adiabatic contraction 
probably overestimates  the dark matter fraction because
 of the hierarchical build-up of the 
galaxies by merging (Ullio, Zhao \& Kamionkowski 2001). 

For black holes with masses $M_{\bullet} > M_{\rm disr}$ the dark matter 
particles will be captured in the same way as the stars. 
A fraction $f_{\rm DM}$ of the total mass captured will be dark matter 
particles. It is interesting to note that for a given density 
distribution the capture rate of collisionless dark matter particles 
is the same as that of collisional  dark matter particles if the 
loss cone can be kept continuously filled. Of course 
if the dark matter cross section is large enough  that the dark matter becomes 
fluid-like then the inner dark matter density distribution may be   
modified (Ostriker 2000).  The dark matter 
particles will also contribute a fraction $f_{\rm DM}$ to 
the hardening of supermassive binary black holes and will 
be ejected from the core in the same manner as stars.

\section{Conclusions} 

The capture of stars and dark matter particles from  orbits 
with sufficiently low angular momentum to pass the event horizon  
contributes significantly to the total mass of black holes 
in the bulges of galaxies if all bulges initially  
had an isothermal  cusp and if no depletion of these orbits occurs. 
The dark matter fraction of the total mass captured is 
20--40 percent for typical CDM-like halos.  
A tight relation between black hole mass and stellar 
velocity dispersion of the form  
$M\propto \sigma_*^{5}$, very similar to the observed relation, arises 
if the black holes in the bulges of galaxies
gain most of their mass by this mechanism. 
The relation is then expected to 
flatten at black holes masses smaller than $10^{8} \msun$ 
where capture and tidal disruption of stars outside  the horizon becomes 
important.  Accretion of gas during the active QSO phase with 
its peak at redshift $z\sim 2-3$ already accounts for a 
significant fraction ($\ge 30 \%$) of the total mass density 
in black holes.  The capture of stars may thus tighten and refine 
the exact slope of a pre-existing rough correlation 
between black hole mass and  bulge properties.   It will thus 
be interesting to see if the tightness of the correlation persists 
when sample sizes get larger. 

In our model bright ellipticals initially had isothermal cusps 
and stars are predominantly captured from the sphere of influence.  
Relaxation of  energy momentum is not required to fill the loss cone
in these isothermal cusps.  The randomization of stellar orbits is
enhanced over the classical two-body relaxation because of processes such as 
the presence of long-lived remnant 
accretion discs left behind from phases of rapid accretion, 
the infall of smaller galaxies predicted by  hierarchical models 
of galaxy formation  and perhaps also the sinking of massive 
black holes left over from a pre-galactic episode of star formation.

The presence of an isothermal cusp and the repopulation of low-angular 
momentum orbits will considerably increase the 
disruption rate of  stars  by supermassive black holes in the 
bulges of bright galaxies at moderate redshifts. Searches for 
high-reshift supernovae may thus detect disruption flares as 
frequently as a few percent of the supernovae rate. The presence 
of an isothermal cusp and efficient loss cone refilling  will 
also accelerate the hardening  of supermassive binary black 
holes so that they will generally reach the separation where 
they can spiral together within a Hubble time due to the emission 
of gravitational waves. The latter strongly supports the idea that 
the cores of elliptical galaxies  have formed very recently due 
to ejection of stars by supermassive binary black holes. 

\section{Acknowledgements} 

We thank Jerry Ostriker for helpful discussions and comments on the manuscript.

\end{document}